\documentstyle[twoside,fleqn,espcrc2,epsfig]{article}

\newcommand{\AmS}{{\protect\the\textfont2
  A\kern-.1667em\lower.5ex\hbox{M}\kern-.125emS}}

\title{The string tensions of the SU(3) representations}

\author{Sedigheh Deldar\address{Department of Physics, 
        Washington University, St. Louis, MO 63130, USA \\ }}
       
\begin{document}

\begin{abstract}
I report on the status of a computation of fundamental and some higher 
representation string tensions in pure gauge SU(3). An $O(a^2)$ tadpole 
improved action and an anisotropic lattice are used. At present, the 
static quark potentials and the string tensions are calculated by measuring 
Wilson loops on an $8^3 \times 24$ lattice. Wilson loops for higher
representations are measured in terms of 
Wilson loops of the fundamental representation. At the small and 
intermediate distances available, rough agreement with Casimir scaling is 
observed, and no color screening for the $8$ representation is seen.
 
\end{abstract}

\maketitle

\section{INTRODUCTION}

According to the theory of confinement, the potential energy of a pair
of heavy static quarks should increase linearly as a string or tube of 
flux is formed between them. For confined representations, one expects
to see a Coulombic plus a linear term for the potential,
$V(r) \simeq -A/r + Kr +C $ , where $K$ is the string tension and 
$r$ is the quark separation. For screened representations, $V(r)$ should 
level off at very large $r$, but one still expects the confined form for 
intermediate $r$ with approximate Casimir scaling of the string tension.
Color screening must occur for adjoint quarks at sufficiently large 
separation, but it is very difficult to observe in numerical 
simulations, at least for zero temperature. The string tensions in
various representations are basic properties of the pure non-abelian 
gauge theory. To be viable, any theory of confinement (monopole 
condensation, center vortices,...) should thus be expected to reproduce 
the string tension in all representations. This point was emphasized by
Greensite at Lat96 \cite{Debb97}. 

The formation of a flux tube and the establishment of a linear potential
between adjoint quarks at non-zero temperature have been shown by many 
numerical experiments in SU(2)\cite{Bern82}. 
Also there are some studies for the existence of the string tension for
the adjoint and a few other non-fundamental representations in SU(3) at 
$T \neq 0$ \cite{Fabe88}.
N. A. Campbell {\it et al.}\cite{Camp86} reported the adjoint string tension 
for SU(3) at low temperature as well.
In this work with an improved action I attempt to determine precisely the
force between static quarks in various representations. So far I have 
been able to show that the string tension at intermediate distances 
is representation dependent and roughly proportion to the quadratic 
Casimir number of the representation.

\section{CALCULATIONS}

The string tension may be found by measuring the Wilson loops and looking
for the area law fall-off for large $t$, $W(r,t)\simeq \exp[-V(r)t]$ where
$W(r,t)$ is the Wilson loop as a function of $r$, the 
spatial separation of the quark, and the propagation time $t$, and $V(r)$
is the gauge field energy associated with static quark-antiquark source.
The interquark energy for large separation grows linearly, so one can 
write $V(r)\simeq -A/r + Kr + C$ for large $r$, with $K$ the string 
tension. In my calculations, I have measured the Wilson loops and found 
the string tension by fitting the data to the above equation. 

Direct measuring of Wilson loops for higher representations by 
multiplying the large matrices is not feasible considering the computer 
memory and the running time. One may expand the trace of 
Wilson loops for higher representations in terms of Wilson loops, $U$, in
the fundamental representation. The higher representation states are 
defined by the tensor product method. Let $W$ be 
the higher representation counterpart to fundamental $U$ for $6, 8, 10, 
15s (symmetric), 15a (antisymmetric),$ or $27$, then: 

\begin{equation}
6:~tr(W) = 1/2~ [~(trU)^2 + trU^2)~ ]
\end{equation}
\begin{equation}
8:~tr(W) = trU^\star trU -1 
\end{equation}
\begin{eqnarray}
10: ~tr(W) = 1/6~[~(trU)^3 + 2(trU^3) \nonumber \\
      + 3trUtrU^2~ ]
\end{eqnarray}
\begin{eqnarray}
15s:~tr(W) = 1/24~[~ (trU)^4 + 6(trU)^2trU^2  \nonumber \\
              +8trU(trU^3) 
              +3(trU^2)^2 + 6trU^4 ~] 
\end{eqnarray}
\begin{eqnarray}
15a:~tr(W) = 1/2trU^\star~[~(trU)^2+ \nonumber \\
               trU^2]-trU
\end{eqnarray}
\begin{eqnarray}
27:~tr(W)=1/4[trU^2+(trU)^2]~[(tr(U^\star)^2) \nonumber \\
   +(trU^\star)^2~]-trUtrU^\star
\end{eqnarray}

\section{SIMULATIONS}

The measurements have been done on an $8^3\times24$ anisotropic lattice
with $a_{s}/a_{t}=3$, where $a_{s}$ and $a_{t}$ are the spatial and 
temporal spacing, respectively. The improved action used for the 
calculations has the form \cite{Morn97}:
\begin{eqnarray}
S = \beta{\{\frac{5}{3}\frac{\Omega_{sp}}{\xi u_{s}^4}+ 
\frac{4}{3} \frac{\xi\Omega_{tp}}{u_{s}^2u_{t}^2}- 
\frac{1}{12}\frac{\Omega_{sr}}{\xi u_{s}^6}-
\frac{1}{12}\frac{\xi\Omega_{str}}{u_{s}^4u_{s}^2 }}\} 
\end{eqnarray}
where $\beta=6/g^2$, $g$ is the QCD coupling, and $\xi$ is the aspect ratio
($\xi=a_{s}/a_{t}$ at tree level in perturbation theory). $\Omega_{sp}$
and $\Omega_{tp}$ include the sum over spatial and temporal plaquettes;
$\Omega_{sr}$ and $\Omega_{str}$ include the sum over $2\times1$ spatial
rectangular and short temporal rectangular (one temporal and two spatial 
links), respectively. For $a_{t} \ll a_{s}$ the discretisation error of 
this action is $O(a_{s}^4,a_{t}^2,a_{t}a_{s}^2)$. The coefficients are 
determined using tree level perturbation theory and tadpole improvement
\cite{Lepa93}. (The spatial mean link, $u_{s}$ is given by $\langle\frac{1}{3}ReTrP_{ss'}\rangle ^\frac{1}{4}$, where $P_{ss'}$ denotes the spatial plaquette. When
$a_{t} \ll a_{s}$, $u_{t}$, the temporal mean link can be fixed to 
$u_{t}=1$, since its value in perturbation theory differs by unity by 
$O(\frac{a_{t}^2}{a_{s}^2})$.)

To minimize the excited state contamination in the correlation functions,
the spatial links are smeared. In the smearing procedure each spatial 
link is replaced by itself plus a sum of its four neighboring
spatial staples times a smearing factor $\lambda$ \cite{Alba87}.
Projection back to SU(3) after smearing or averaging over different 
paths in Wilson loops is necessary, since I want to use eqns. $ 1-6 $
in which $U's$ should be SU(3) matrices. 
For the same reason I am not able to do thermal 
averaging, since again it would take links out of SU(3). (Thermal 
averaging is the replacement of a timelike link by its average with fixed neighbors, which is normally useful to increase statistics.)

I have used MILC Code as a platform for the simulations and my code has 
been run on a Dec Alpha and the Origin2000 supercomputer at NCSA 
(single node jobs).
The potential for on- and off-axis points is calculated by measuring 
Wilson loops. Fig 1 shows a typical plot for representation $8$. $Q$, the 
confidence level of the fit, is calculated by measuring the covariance 
matrix evaluated by the jack knife method. Fig. 2 shows $a_{t}v(r)$ versus 
$r$ for the fundamental, $6$ and $8$ representations. The data have been 
fitted to a Coulombic plus linear term. Only the on-axis points
are used in the fits. 
Without considering systematic errors (in particular, the small violation
of rotational invariance), the confidence levels of the fits
are not very good. 
Results are shown in Table 1. There is rough 
agreement between $\frac{K}{K_{f}}$ or $\frac{A}{A_{f}}$  with 
$\frac{C}{C_{f}}$ as predicted \cite{Ambj84}. Also $\frac{K_{8}}{K_{f}}$ 
calculated by this work 
is in agreement with $2.2(4)$ reported by N. A. Campbell {\it et al.} 
\cite{Camp86}. For $\beta=2.4$, I have found $a_{s} \simeq .28 f$ by 
comparing the fundamental string tension with the phenomenological value
$K=420$ MeV.

\begin{figure}[htb]
\vspace{25pt}
\epsfxsize=1. \hsize
\epsffile{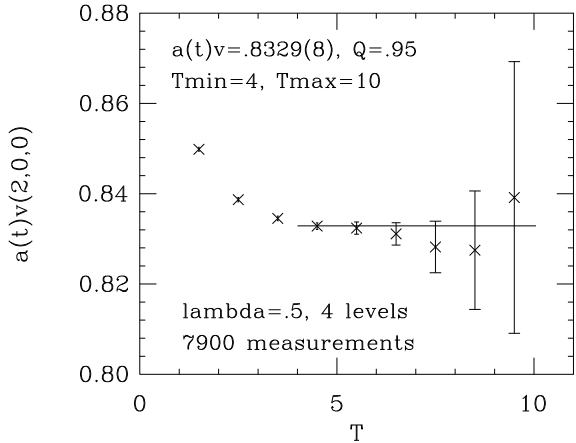}
\vspace{-50pt}
\caption{Effective mass plot for representation 8. Tmin and Tmax show 
the range of the fit. $\lambda$, the smearing factor, and the smearing 
levels are determined by some low statistics runs for short, intermediate
and long distances. }
\label{fig:largenenough}
\vspace{-15pt}
\end{figure}

\begin{figure}[htb]
\vspace{25pt}
\epsfxsize=1. \hsize
\epsffile{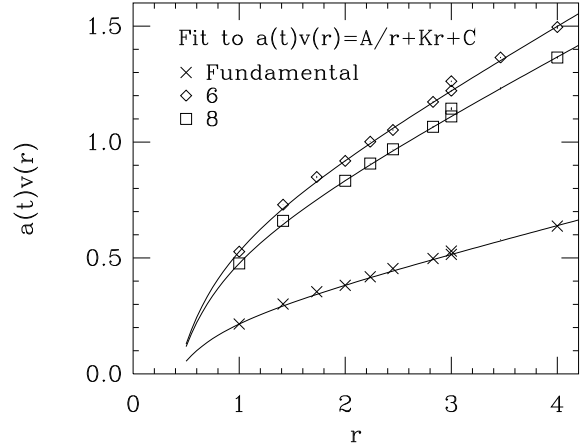}
\vspace{-50pt}
\caption{Potentials for the fundamental, 6, and 8 representations.
Only on axis points are used in the fits. The fits are based
on 7200, 7200 and 7900 measurements for fundamental, 6 and 8 
representations respectively. Rough agreement with
Casimir scaling is observed, and no color screening for 8 representation 
is seen.}
\label{fig:toosmall}
\vspace{-15pt}
\end{figure}

\section{CONCLUSIONS}

By measuring the Wilson loops for the fundamental and some higher 
representations ($6$ and $8$), I have shown that the string tensions 
exist and their values are different for each representation and are in 
rough agreement with Casimir scaling. At the small and intermediate 
distances available, no color screening is observed. In the future, I plan
to finish the analysis for representations $10, 15s, 15a,$ and $27$ by 
collecting more data. To take the continuum limit I also need weaker and 
stronger coupling results. Finally, I plan to look at larger lattices in 
the hope of seeing evidence of the approach to the asymptotic form of 
the potential.

\begin{table}[htb]
\vspace{-15pt}
\setlength{\tabcolsep}{.5pc}
\newlength{\digitwidth} \settowidth{\digitwidth}{\rm 0}
\catcode`?=\active \def?{\kern\digitwidth}

\caption{Parameters of the potentials as a function of representation. K
is the string tension, A is the Coulombic coefficient term, C is the 
Casimir number and ``f`` stands for the fundamental representation. The 
errors shown are statistical only.}
\begin{tabular}{lcccc}
\hline
$Repn.$  &{$K$} & {$\frac{K}{K_{f}}$} & {$\frac{A}{A_{f}}$} & {$\frac{C}{C_{f}}$} \\
\hline
$3$ & $ .3480(6)$     & -         & - & -   \\
$6$ & $ .7688(9)$  & $2.209(5)$       & 2.65(2)        &$2.5$   \\
$8$ & $.710(1)$    & $2.040(5)$   &$2.35(1)$  &$2.25$ \\
\hline
\end{tabular}
\vspace{-25pt}
\end{table}

\section{ACKNOWLEDGEMENT}

I would like to thank my advisor Claude Bernard for his great help in 
this work. Also I wish to thank J. Mandula, U. Heller and C. Morningstar
for their patience in answering my questions and R. Sugar for helping me
to begin computing on the Origin2000.

\end{document}